\documentclass[sigconf]{acmart} 
\usepackage{booktabs,makecell,multirow}
\AtBeginDocument{%
  \providecommand\BibTeX{{%
    \normalfont B\kern-0.5em{\scshape i\kern-0.25em b}\kern-0.8em\TeX}}}

\setcopyright{acmcopyright}
\copyrightyear{2018}
\acmYear{2018}
\acmDOI{10.1145/1122445.1122456}

\copyrightyear{2024}
\acmYear{2024}
\setcopyright{acmlicensed}\acmConference[CVIPPR 2024]{2024 2nd Asia Conference on Computer Vision, Image Processing and Pattern Recognition}{April 26--28, 2024}{Xiamen, China}
\acmBooktitle{2024 2nd Asia Conference on Computer Vision, Image Processing and Pattern Recognition (CVIPPR 2024), April 26--28, 2024, Xiamen, China}
\acmDOI{10.1145/3663976.3664242}
\acmISBN{979-8-4007-1660-7/24/04}

\begin{document}

\title{Contrastive Learning with Spectrum Information Augmentation in Abnormal Sound Detection}

\author{Xinxin Meng}
\authornote{Both authors contributed equally to this research.\\
This work was supported in part by the National Natural Science Foundation of China under Grant 62366051, and in part by the State Grid Xinjiang Electric Power Company and Xinjiang Siji Information Technology Co., Ltd. under Grant SGITXX00ZHXX2200262.}
\email{1731224941@foxmail.com}
\affiliation{%
  \institution{Information and Communication Company, State Grid Xinjiang Electronic Power Company}
  \streetaddress{123 Jianshe Road, Tianshan District}
  \city{Urumqi}
  \state{The Xinjiang Uyghur Autonomous Region}
  \country{China}
  \postcode{830017}
}

\author{Jiangtao Guo}
\email{gjt933@foxmail.com}
\affiliation{%
  \institution{Information and Communication Company, State Grid Xinjiang Electronic Power Company}
  \streetaddress{123 Jianshe Road, Tianshan District}
  \city{Urumqi}
  \state{The Xinjiang Uyghur Autonomous Region}
  \country{China}
  \postcode{830017}
}

\author{Yunxiang Zhang}
\authornotemark[1]
\email{zhangyunxiang@stu.xju.edu.cn}
\affiliation{%
  \institution{Xinjiang university}
  \streetaddress{666 Shengli Road, Tianshan District}
  \city{Urumqi}
  \state{The Xinjiang Uyghur Autonomous Region}
  \country{China}
  \postcode{830017}
}

\author{Shun Huang}
\email{huangswt@stu.xju.edu.cn}
\affiliation{%
  \institution{Xinjiang university}
  \streetaddress{666 Shengli Road, Tianshan District}
  \city{Urumqi}
  \state{The Xinjiang Uyghur Autonomous Region}
  \country{China}
  \postcode{830017}
}
\renewcommand{\shortauthors}{Trovato and Tobin, et al.}

\begin{abstract}
  The outlier exposure method is an effective approach to address the unsupervised anomaly sound detection problem. The key focus of this method is how to make the model learn the distribution space of normal data. Based on biological perception and data analysis, it is found that anomalous audio and noise often have higher frequencies. Therefore, we propose a data augmentation method for high-frequency information in contrastive learning. This enables the model to pay more attention to the low-frequency information of the audio, which represents the normal operational mode of the machine. We evaluated the proposed method on the DCASE 2020 Task 2. The results showed that our method outperformed other contrastive learning methods used on this dataset. We also evaluated the generalizability of our method on the DCASE 2022 Task 2 dataset.
\end{abstract}



\maketitle

\section{Introduction}
Due to the rapid development of artificial intelligence technology, anomaly detection algorithms based on machine learning techniques utilizing image features have become essential tools for monitoring industrial equipment operations and detecting machine faults at an early stage \cite{STAAR2019484,9054344}. However, the complex structure of machines and the large number of components mean that many endogenous faults may not be visible from the exterior of the machine. In contrast, audio data offers several advantages over image data, including resilience to weather conditions and no blind spots\cite{10096813,10095736}. Therefore, employing machine learning techniques for anomaly detection based on audio signals is a valuable addition to the application of artificial intelligence in machine fault diagnosis. Machine Abnormal Sound Detection (ASD) faces several challenges, including:
\begin{itemize}
\item \textbf{Data imbalance problem:} The amount of data available for anomalous sounds is often much less than that for normal sounds. 
\item \textbf{The complexity of sound:} Sound signals are complex and contain multiple frequency and amplitude components, making the detection of anomalous sounds challenging. To overcome this challenge, effective feature extraction and dimensionality reduction techniques are required to simplify the sound data and enable accurate detection of anomalies.
\item \textbf{Generalization capability of models:} Anomalous sounds can be diverse, and as a result, a detection model may only be effective for a specific type of anomalous sound or dataset. 
\end{itemize}  \par
Outlier detection is a common approach to deal with several of the problems mentioned above.
By minimizing the contrastive loss, it is possible to effectively extract the normal audio pattern of machine operation and reduce the negative impact of data imbalance problem, improve the generalization ability of the model. This approach has been shown to improve the anomaly detection ability of the model \cite{9746207,10095568,9878474,8237871,9782500}. How to construct positive and negative sample pairs is the most critical design principle, GeCo \cite{10095568} uses a Predictive AutoEncoder (PAE) equipped with self-attention as a generative model to generate negative samples. This approach involves training the feature extractor using pairs of original samples as positive and negative samples to improve the accuracy. Hojjati et al.\cite{9746207} used various audio enhancement techniques such as Pitch Shift, Time Stretch, White Noise Injection, and Time Frequency Masking to perform data augmentation. By constructing positive and negative sample pairs using these techniques, they were able to improve the accuracy of the ASD model.\par
\begin{figure}
\includegraphics[width=\linewidth]{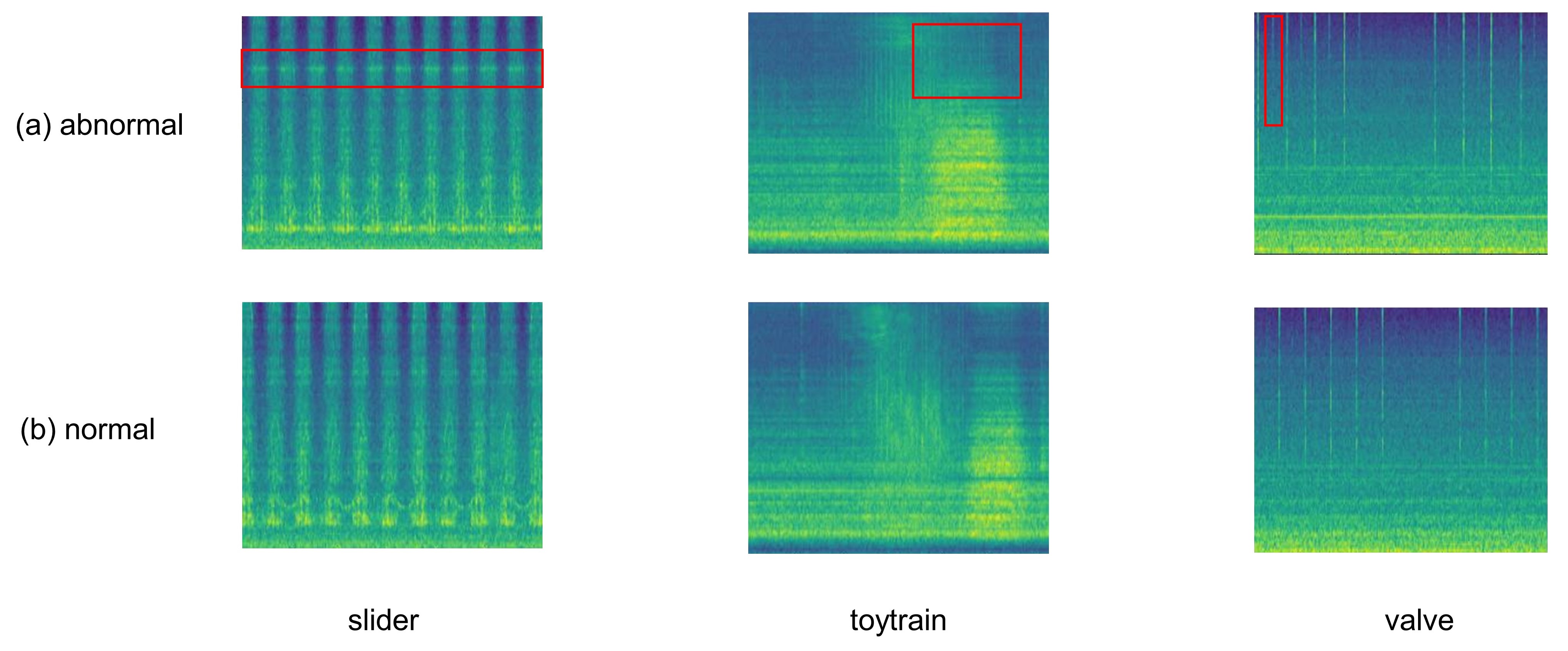}
\caption{The figure displays the spectrograms of three different machine types, with the first row representing abnormal audio spectrograms and the second row representing normal operating audio spectrograms. In each spectrogram, the horizontal axis represents the time domain, and the vertical axis represents the frequency domain. It is evident from these spectrograms that anomalies tend to occur predominantly in high-frequency features (indicated by red boxes).} 
\label{fig1}
\end{figure}

However, the above methods are not based on the characteristics of the data itself. By combining the physical properties of audio recordings\cite{nunes2021anomalous}, we observed the spectrograms of machine audio and found that anomalies and noises tend to occur predominantly in high-frequency ranges, as depicted in (see Fig.~\ref{fig1}). The focus of our attention should be on the normal audio pattern of machine operation. However, neither the generative model nor the basic audio enhancement methods can effectively generate two audios with a large contrast in high-frequency information. This limitation results in the model paying more attention to the normal audio pattern of the audio features during training. To address this issue, we propose a novel data augmentation method based on spectral information augmentation in the field of machine abnormal sound detection, and combine this method with contrastive learning to extract normal audio pattern in audio signals.
This method employs one audio recording for data augmentation, generating two audio recordings with large contrast in high-frequency information from two different perspectives. By doing so, the model can focus more on the normal audio pattern in the audio during the training process and extract the normal pattern of audio operation by minimizing the contrastive loss. Indeed, the normal data in the test data should follow the same pattern as the normal data in the training data. By ensuring the consistency of the normal data distribution of the machine, the model can more accurately assign high anomaly scores to abnormal samples during inference.\par

We evaluate the effectiveness of our proposed method on the DCASE 2020 Task2 evaluation dataset. In DCASE 2020 task 2, we achieved a significantly higher performance with an AUC(Area Under the Curve) of 93.83\% and pAUC(partial Area Under the Curve) of 87.6\%, outperforming the top-ranked system with an AUC of 90.47\% and pAUC of 83.61\%. Additionally, our proposed method outperformed other studies that used contrastive learning methods on this dataset. In the experiments on the evaluation dataset of task 2 in DCASE 2022, we demonstrated that our method has some generalization ability in domain generalization tasks.

\section{PROPOSED METHOD}
\begin{figure*}[htb]
\centering
\centerline{\includegraphics[width=\textwidth]{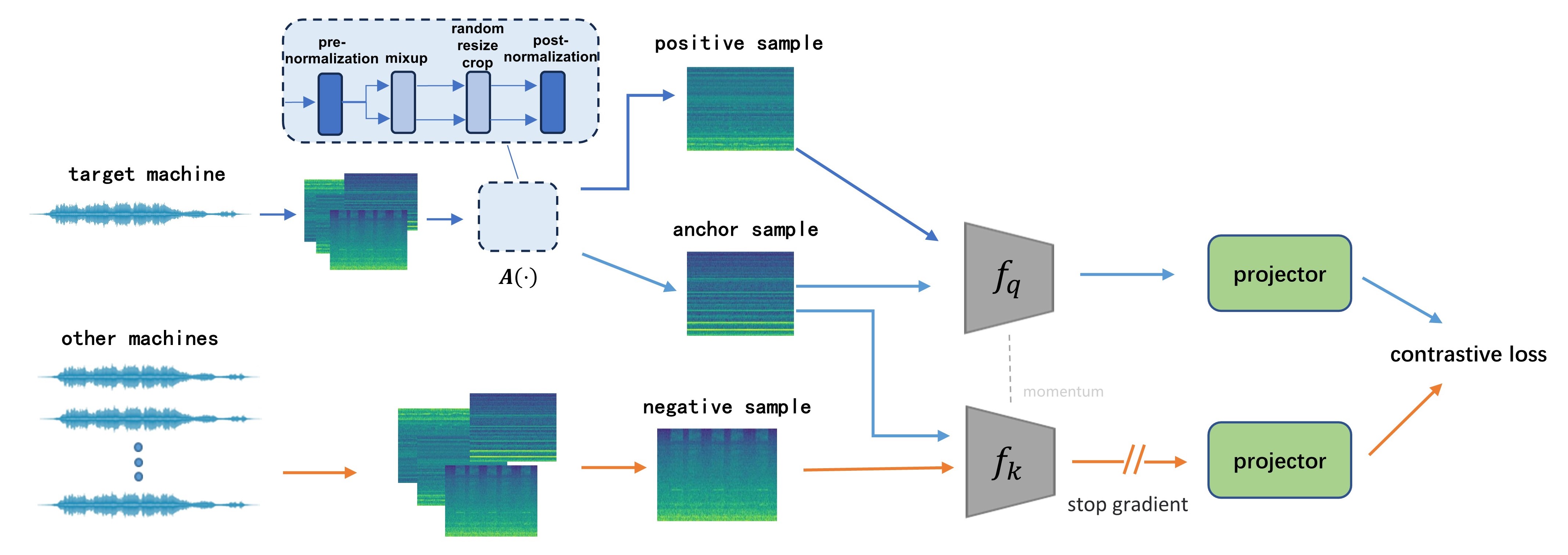}}
\caption{\textbf{The overall framework of our proposed method.} We generate samples of different domains but of the same class. $f_q$ and $f_q$ refers to encoder. $A(\cdot)$ refers to different data augmentation methods. 
These positive and negative sample pairs are then fed into the embedding model $f_q$ for training. The encoder $f_q$ performs normal gradient updates while $f_k$ does not perform gradient updates but employs the momentum update method instead. We extracted The normal pattern of machine audio more accurately by minimizing the contrastive loss and training the model to reduce the intra-class distance while increasing the inter-class distance. This approach helps to reduce the distance between source domains and learn cross-domain invariant representations present in the data, ultimately enhancing the model's generalization capabilities across diverse acoustic environments.}
\label{Fig.1}
\end{figure*}

In this section, we will provide a detailed introduction to our approach. Compared to the study by Hojjati et al. \cite{hojjati2022self}, we utilized a more comprehensive dataset for testing and a simpler data augmentation method, yielding improved results on the MIMII dataset \cite{Purohit_DCASE2019_01}. The MIMII dataset contains data from four machine types fan, pump, slide, and valve. The overall framework is depicted in Fig.\ref{Fig.1}.
Based on the data characteristics of the ASD task, we propose a contrastive learning framework with large high-frequency feature contrast. This approach takes advantage of the official datasets provided by DCASE 2020 Task 2, which include data from multiple sections for each machine type, each with a different distribution due to varying machine operating parameters. Within the same section, the data distribution is close to each other. To increase the diversity of the data, we performed direct data augmentation on samples from each section, generating positive and anchor samples, while also including data from other sections of the same machine type as negative samples. For DCASE 2022 Task2, we adopt the method of pre-training and fine-tuning the downstream classification model to detect abnormal data. 
Our approach as the upstream model, using Audioset \cite{gemmeke2017audio} for pre-training. We use instance discrimination \cite{wu2018unsupervised} as the pretext task for pre-training the model, where each audio record is regarded as one class. We generate two records with large differences in high-frequency feature by using different perspectives data enhancement for the same audio record.\par

We employ the following transformations to generate two spectrograms with large contrast in high-frequency information:
\begin{itemize}
\item \textbf{Pre-Normalization:} The data is normalized by using the mean and standard deviation of the training samples. This approach helps to stabilize the calculations in the system.
\item \textbf{Mixup for high-frequency information:} We employ the basic mixup calculation as an augmentation technique, while was originally designed for mixing both features and labels. However, in our case, we apply it exclusively to audio features only (because of the absence of labels). Considering that audio is log-scaled, we first convert the input to a linear scale before performing the mixup calculation, and then revert it back to a log-scale. We refer to this process as Log-mixup-exp. Specifically, the Log-mixup-exp of the $i$th input $x_i$ is: 
\begin{equation}
\overset{\sim}{x}_i = \log((1 - \lambda)\exp(x_i) + \lambda\exp(x_k)),
\end{equation}
where $x_k$ is a mixing counterpart, and mixing ratio $\lambda$ is sampled from uniform distribution $U(0.0, 
\alpha)$, as opposed to the beta distribution utilized in the original mixup algorithm. Furthermore, $\alpha$ is a mixing ratio hyper-parameter that controls the contrast degree between the resulting mixed outputs, and we set $\alpha$ to 0.4. The mixing counterpart $x_k$ is randomly chosen from a memory bank, which is a FIFO queue that stores past inputs.\par

The Mixup block takes the normalized log-mel spectrogram audio as input and mixes small proportions of the past randomly selected input audio. In audio recording, we use background sound for mixing since acoustic events predominantly occur in foreground sounds. Consequently, this generates contrast in the background sound of the mixup output pairs, thereby promoting the learning of invariant foreground acoustic event representations.
\item \textbf{Random Resize Crop (RRC):} We employ RRC (Random Resized Crop) as an approximation of pitch change and time extension in the log-mel spectrogram, with the aim of learning meaningful representations within audio recordings.
\item \textbf{Post-Normalization:} Since the previous enhancement operation may have induced data shift, we utilize a post-normalization operation to rectify the drift, ensuring that the final output views conform to a standard normal distribution, $~N(0,1)$.

\end{itemize}
\par
After preparing the data, we select samples from different domains within the same class to construct positive sample pairs \cite{niizumi2021byol}. One sample serves as an anchor, while the other functions as a positive sample. Subsequently, we combine the anchor sample with data from other classes to form negative sample pairs. We feed these sample pairs into the same encoder for training. Notably, the key network does not perform gradient updates but instead adopts a momentum update strategy \cite{he2020momentum}. Formally, we denote the gradient parameters of $f_k$ as $\theta_k$ and those of $f_q$ as $\theta_q$, we update $\theta_k$ by: 
\begin{equation}
\theta_k \leftarrow m\theta_k + (1-m)\theta_q
\end{equation}
\noindent in this context, $m \in[0,1)$ is a momentum coefficient. Only the parameters $\theta_q$ are updated through back-propagation. Here we set $m$ to 0.99 with the same parameter settings for both datasets. The objective of contrastive learning is to bring the anchor and positive groups closer together while pushing the anchor and negative groups further apart. Since we select positive and negative sample pairs based on class-condition, the intra-class distances continually decrease, and the inter-class distances are increase during training. We calculate the loss using the embedding output of the two encoders. By minimizing the contrastive loss, we gradually eliminate the distance between different devices of the same class of machines.

\begin{table*}[htbp]
\caption{\centering AUC(\%) and pAUC(\%) performance comparison of different machines in DCASE 2020 Task 2 evaluation set.}
\resizebox{\textwidth}{18mm}
{
\begin{tabular}{ccccccccccccccc} 
\toprule[2pt]
\multirow{2}*{Methods}
&\multicolumn{2}{c}{Fan} &\multicolumn{2}{c}{Pump}  &\multicolumn{2}{c}{Slider} &\multicolumn{2}{c}{Valve} &\multicolumn{2}{c}{ToyCar} &\multicolumn{2}{c}{ToyConveyor} &\multicolumn{2}{c}{Average} \\
\cmidrule(r){2-3} 
\cmidrule(r){4-5} 
\cmidrule(r){6-7} 
\cmidrule(r){8-9} 
\cmidrule(r){10-11} 
\cmidrule(r){12-13} 
\cmidrule(r){14-15}
 & AUC & pAUC & AUC & pAUC & AUC & pAUC & AUC & pAUC & AUC & pAUC & AUC & pAUC & AUC & pAUC \\ 
\midrule 
Glow-Aff \cite{dohi2021flow} & 74.90 & 65.30 & 83.40 & 73.80 & 94.60 & 82.80 & 91.40 & 75.00 & 92.20 & 84.10 & 71.50 & 59.00 & 85.20 & 73.90 \\
ESResNeXt \cite{guzhov2021esresne} & 85.52 & 77.29 & 92.88 & 85.73 & 91.08 & 79.82 & 86.01 & 86.16 & 75.1 & 65.44 & 88.77 & 86.94 & 86.56 & 80.23 \\
M-A-G-IDcl \cite{Giri2020} & 83.06 & 79.55 & 87.04 & 79.52 & 97.43 & 88.91 & 95.27 & 90.23 & 79.1 & 64.02 &\textbf{99.07} & \textbf{96.20} & 90.16 & 83.07 \\
STgram-MFN \cite{liu2022anomalous} & 88.09 & 88.06 & 87.31 & 85.42 & \textbf{99.29} & \textbf{96.93} & 91.97 & 89.47 & 94.32 & 90.17 & 82.87 & 76.48 & 91.97 & \textbf{89.47}\\
SSAAD \cite{hojjati2022self} & 80.11 & N/A & 70.12 & N/A & 77.43 & N/A & 84.17 & N/A & N/A & N/A & N/A & N/A & N/A & N/A \\
GeCo \cite{10095568} & \textbf{92.73} & 85.19 & \textbf{93.09} & \textbf{86.89} & 98.61 & 95.26 & \textbf{99.06} & \textbf{95.52} & \textbf{96.62} & 89.33 & 74.69 & 65.82 & 92.47 & 86.34\\
Top-1 \cite{Giri2020} & 82.39 & 78.23 & 87.64 & 82.37 & 97.28 & 88.03 & 98.46 & 94.87 & 95.57 & \textbf{91.54} & 81.46 & 66.62 & 90.47 & 83.61\\
\midrule 
Ours &91.34 & \textbf{90.05} & 91.97 & 78.17 & 96.15 & 89.82 & 92.18 & 86.9 & 92.73 & 85.07 & 98.82 & 95.59 & \textbf{93.83} & 87.6  \\
\bottomrule[2pt]     
\end{tabular}}
\label{table:1}
\end{table*}

\section{Experiments}
\subsection{Datasets}
We use Audioset \cite{gemmeke2017audio}, a large-scale unsupervised pre-training dataset provided by Google in 2017, for pre-training. This dataset contains 632 audio categories and 2,084320 manually labeled sound clips (including 527 labels), each with a length of 10-seconds. The tags cover human voice, environment sound, machine sound and other fields. This dataset has been used multiple times in large-scale unsupervised audio pre-training tasks \cite{niizumi2021byol,saeed2021contrastive}. The dataset for our detection is the official dataset published in Task 2 of DCASE 2020 and DCASE 2022 \cite{dohi2022mimii,Harada2021}. The data in DCASE 2022 includes normal/abnormal operating sounds of seven types of machines, with each recording being a single-channel 10-second audio clip. Each machine type provides data in multiple different domains for training and validation in the training set and validation set. The domain to which the test data belongs is unseen. The DCASE 2020 dataset consists of two datasets, ToyADMOS \cite{Harada2021} and MIMII \cite{Purohit_DCASE2019_01}, which include six types of operation sounds from normal/abnormal toy/real machines, and the other parameters are the same as the dataset of DCASE 2022 Task 2. Both datasets provide two parts: training set and evaluation set.

\subsection{Implementation}
In DCASE 2020, we trained one model for per ID(as section) for each machine type. The training data consisted of original audio waveforms with a duration of 10 seconds. We used log-Mel spectrogram as input features, with a frame size of 1024 and each of sample with an overlapping 50\% in the spectrogram. The number of Mel filter banks was set to 128 (i.e., W = 1024, H = 512 and M = 128). After feature extraction, the dimension of each log-Mel spectrogram was 128x313. We used the Adam optimizer with a learning rate of 0.0009 and trained each model for 150 epochs with a batch size of 32.\par

In the DCASE 2022 experiment, we adopted the method of pre-training and fine-tuning a downstream classification model. In the pre-training stage, we employed a queue structure of length $n$ to store $ \{k_1,k_2,\ldots,k_n\}$(considered negative keys). By doing so, a large number of negative sample dictionaries were maintained. Contrastive learning benefits from comparing more negative samples with the same positive sample pair. For the encoder structure, we utilized ResNet50 \cite{he2016deep}. In the downstream task, we employ the idea of outlier detection and construct individual spatial planes for each machine type using a classification approach. During testing, the distance between the test vector and the samples in the space is computed, and the farther the distance, the higher the degree of abnormality. 
The log-Mel spectrogram is used as the input features, and the feature extractor in the query network is utilized to load the classification model and fine-tune the whole network structure. In the downstream classification model, we use two CNN-based network structures namely MobilenetV2 \cite{sandler2018mobilenetv2} and MobileFacenet \cite{chen2018mobilefacenets}, to evaluate the effectiveness of our proposed method. During testing, the distance between the test vector and the samples in the space is computed, and the farther the distance, the higher the degree of abnormality.

After the encoder output, we calculate the similarity of the two groups of vectors using dot product, and utilized InfoNCE \cite{oord2018representation} as the loss function in the pre-training stage. The calculation method is as follows: 
\begin{equation}
\mathcal{L}_q=-log\frac{\exp(q\cdot k_+/\tau)}{\sum\limits_{i=0}\limits^{n}\exp(q\cdot k_i/\tau)},
\end{equation} 
\noindent where $\tau$ is a temperature hyper-parameter per \cite{wu2018unsupervised}.

\section{Results AND Discussion}
We evaluated the performance of our model using the Area Under the receiver operating characteristic Curve (AUC) and the partial-AUC (pAUC) \cite{baijless}. The pAUC is calculated as the AUC over a low False-Positive-Rate (FPR) range $[0, p]$ and $p$ = 0.1. Table \ref{table:1} shows the results on the dataset in DCASE 2020 Task 2. Our proposed method outperforms other approaches \cite{hojjati2022self,10095568} that utilize contrastive learning for extracting representations in the field of machine anomaly sound detection. The table also shows that the proposed method achieved 93.83\% and 87.60\% in terms of the AUC and pAUC, respectively, giving a 3.36\% and 3.99\% improvement over the top-ranked model in the challenge. After examining the experimental results, it is evident that our proposed method has a remarkable ability to improve model domain generalization ability. In DCASE 2020, overall, our method outperforms other methods in terms of average evaluation metrics across all machine types. From the perspective of a single machine type, our method is either the best or the second best, providing further evidence of the generalization ability of our model. Moreover, our method is effective across all machine types.



Table \ref{table:2} demonstrates the effectiveness of our proposed method on the dataset in DCASE 2022 Task 2. The results shows that our proposed method improves the generalization ability across all machine types. According to the table, our proposed method achieved 74.82\% and 66.69\% in terms of the AUC in the source domain and target domain.  These results indicate a 26.62\% and 16.83\% improvement over the baseline model.The significant improvements in the source and target domains prove that our method is also competitive in the domain generalization ability, which significantly improves the generalization ability of the model. The powerful ability of CLDG to generalize in the face of multiple unknown domains is demonstrated. The basic model used as a baseline model was provided by the official, and we did not adjust the training parameters of the original model When fine-tuning the classification model. It should be noted that many of the top-performing systems in the competition were ensemble models, which we did not include in our analysis. \par
We utilized the entire Audioset for pre-training, which contains almost all kinds of sounds worldwide, but this may affect the final results when applied to downstream tasks. Therefore, our next experimental direction is to incorporate more machine audio data during performing pre-training.

\begin{table*}[htbp]
\caption{\centering AUC(source)(\%) and AUC(target)(\%) performance comparison of different types of machines in DCASE 2022 Task 2 evaluation set, where MN refers to MobilenetV2, P refers to pre-trained, MFN refers to MobileFacenet.}
\resizebox{\textwidth}{12.5mm}{
\begin{tabular}{ccccccccccccccccc} 
\toprule[2pt]
\multirow{2}*{Methods}
&\multicolumn{2}{c}{Fan} &\multicolumn{2}{c}{Bearing}  &\multicolumn{2}{c}{Slider} &\multicolumn{2}{c}{ToyCar} &\multicolumn{2}{c}{ToyTrain} &\multicolumn{2}{c}{Valve} &\multicolumn{2}{c}{Gearbox}    &\multicolumn{2}{c}{Average} \\
\cmidrule(r){2-3} 
\cmidrule(r){4-5} 
\cmidrule(r){6-7} 
\cmidrule(r){8-9} 
\cmidrule(r){10-11} 
\cmidrule(r){12-13} 
\cmidrule(r){14-15}
\cmidrule(r){16-17}
 & AUC(s) & AUC(t) & AUC(s) & AUC(t) & AUC(s) & AUC(t) & AUC(s) & AUC(t) & AUC(s) & AUC(t) & AUC(s) & AUC(t) & AUC(s) & AUC(t) & AUC(s) & AUC(t) \\ 
\midrule 
\textbf{MN(baseline)}  & 61.64 & 57.56 & 62.43 & 57.89 & 59.21 & 56.49 & 57.00 & 52.46 & 53.34 & 51.58 & 64.84 & 65.26 & 64.70 & 56.53 & 60.45 & 56.82 \\
MN+P  & \textbf{71.66} & 46.93 & 62.60 & \textbf{70.35} & 74.13 & 58.36 & 74.47 & 70.83 & \textbf{62.66} & 58.84 & 75.97 & 76.71 & 74.01 & 66.51 & 67.43 & 59.70 \\
\midrule
MFN & 62.79 & 41.77 & 57.08 & 60.20 & 71.66 & 54.39 & \textbf{76.83} & \textbf{78.90} & 60.36 & \textbf{63.33} & 72.28 & 69.04 & 71.25 & 55.18 & 63.93 & 56.89 \\
MFN+P & 71.52 & 45.09 & \textbf{65.93} & 69.80 & \textbf{82.70} & \textbf{69.30} & 73.35 & 72.16 & 61.80 & 54.51 & \textbf{81.67} & \textbf{80.38} & \textbf{86.78} & \textbf{75.58} & \textbf{74.82} & \textbf{66.69} \\
\bottomrule[2pt]     
\end{tabular}}
\label{table:2}
\end{table*}

\section{Conclusion}
In this paper, we propose a contrastive learning method based on high-frequency information augmentation in the unsupervised machine ASD, which is suitable for machine-generated audio. Experimental results show that our proposed method outperforms the top-ranked system in DCASE 2020 Task 2 and surpasses other methods that apply contrastive learning to ASD tasks. 
In the experiment of DCASE 2022 Task 2, it is verified that our model has certain generalization ability.
The experimental results on both DCASE 2020 Task 2 and DCASE 2022 Task 2 tasks demonstrate the effectiveness of our method in improving the performance of machine ASD, highlighting the potential of our approach in addressing the challenges of domain generalization in machine ASD.

\bibliographystyle{ACM-Reference-Format}
\bibliography{main}


\end{document}